\pdfoutput=1
\documentclass[a4paper, oneside, twocolumn, notitlepage, 10pt]{extarticle_ecoc}
\usepackage{ecoc}
\usepackage{physics}
\usepackage{xcolor}
\usepackage{setspace}
\usepackage{physics}
\usepackage{color, xcolor}
\usepackage{soul}

\hypersetup{
    pdftitle={Experimental Quantification of Layered Error Suppression in Fiber-Interconnected Quantum Data Centers},
    pdfauthor={Seyed Navid Elyasi, Sima Bahrani, Rui Wang, Dimitra Simeonidou, Paolo Monti, Rui Lin},
    pdfsubject={ECOC 2026 Submission},
    pdfkeywords={Distributed Quantum Computing, Quantum Data Centers, Error Mitigation, Entanglement Purification}
}
\begin{document}
\selectlanguage{english}    


\title{Experimental Quantification of Layered Error Suppression in Fiber-Interconnected Quantum Data Centers}%

\author{
    Seyed Navid Elyasi\textsuperscript{(1)}, Sima Bahrani\textsuperscript{(2)},
    Rui Wang\textsuperscript{(2)},
    Dimitra Simeonidou\textsuperscript{(2)}, 
    Paolo Monti\textsuperscript{(1)},
    Rui Lin\textsuperscript{(1)}}
\maketitle                  


\begin{strip}
    \begin{author_descr}

        \textsuperscript{(1)} Department of Electrical Engineering, Chalmers University of Technology, Gothenburg, Sweden
        
        \textsuperscript{(2)} Smart Internet Lab, School of Electrical, Electronic, and Mechanical Engineering, University of Bristol, Bristol
        \textcolor{blue}
        {\uline{elyasi@chalmers.se}}

    \end{author_descr}
\end{strip}

\renewcommand\footnotemark{}
\renewcommand\footnoterule{}


\begin{strip}
    \begin{ecoc_abstract} 
We perform experiments to quantify error suppression in fiber-connected superconducting QPUs using combined error mitigation techniques, demonstrating over 20\% improvement in operational fidelity across interconnected quantum processing units under realistic noise conditions. \textcopyright2026 The Author(s)



    \end{ecoc_abstract}
\end{strip}


\section{Introduction}

\begin{figure*}[b!]
\begin{center}
\includegraphics[width=1\textwidth]{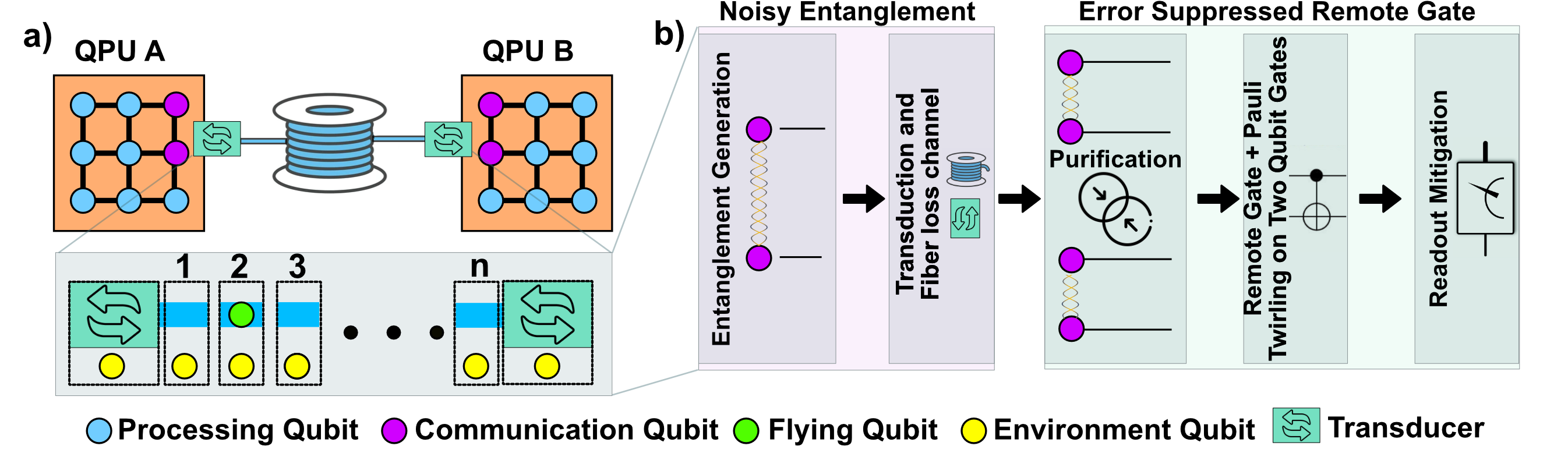}
\end{center}
\caption{(a) Fiber-interconnected superconducting QPUs with a transduction interface, where communication noise is emulated via a quantum collisional model. (b) Protocol layers including entanglement generation, noisy transmission, purification, Pauli-twirled two-qubit gates, and readout mitigation. Colors indicate qubit roles and system components.}
\label{f:Layout}
\end{figure*} 

Quantum computing is advancing rapidly; however, its scalability remains fundamentally limited by the number of qubits that can be reliably integrated on a single chip due to control complexity and cross-talk errors \cite{sebastiano2019,battistel2023,sarovar2020}. To overcome this limitation, quantum data centers (QDCs) have been proposed, where multiple quantum processing units (QPUs) are interconnected to enable distributed quantum computing \cite{liu2024quantum,cacciapuoti2020}. Such architectures extend computational capabilities beyond monolithic processors 
\cite{liu2024quantum, buhrman2003distributed}.

In these systems, QPUs are connected via quantum channels and rely on entanglement to perform nonlocal operations \cite{campbell2024quantum}. A fundamental requirement for universal computation is the realization of high-fidelity remote two-qubit gates, as formalized by the DiVincenzo criteria \cite{divincenzo2000}. Protocols such as cat-state and teleportation-based approaches have been proposed \cite{yimsiriwattana2004,wan2019, campbell2024quantum}. However, their performance is strongly constrained by communication-induced noise \cite{elyasi2025qdc, Elyasi2025QDCTopologies, Elyasi2025FrameworkQDC, bahrani2023analysing}.

While error mitigation and correction techniques are well developed for monolithic processors \cite{egan_2021_fault_tolerant, google_2025_below_threshold}, their experimental evaluation in distributed quantum architectures remains limited, particularly in quantifying their effectiveness under communication-induced noise in interconnected systems. This is particularly true for optically interconnected superconducting platforms, where optical links and microwave-to-optical transduction \cite{mirhosseini2020,rochman2023,campbell2024quantum} degrade entanglement distribution and remote gate fidelity. 

In this paper, we experimentally investigate error suppression for remote gate execution in fiber-interconnected superconducting quantum processors as illustrated in Fig.~\ref{f:Layout}(a). Building on our prior work \cite{elyasi2025qdc}, communication-induced noise is emulated using a quantum collisional model, enabling the controlled injection of transduction and fiber effects within a single device. Experiments are performed on real IBM superconducting hardware under realistic conditions. As shown in Fig.~\ref{f:Layout}(b), we apply a layered error suppression strategy combining entanglement purification to suppress communication-induced noise, Pauli twirling to correct local operation errors, and readout mitigation to overcome noisy measurement during circuit execution.

Our results demonstrate that this layered approach significantly enhances robustness against communication-induced noise, achieving over 20\% fidelity improvement. 
These findings provide practical insights toward scalable distributed quantum computing and the realization of quantum data center architectures.

\section{System and Architecture}

\begin{figure*}[t!]
\begin{center}
\includegraphics[width=1\textwidth]{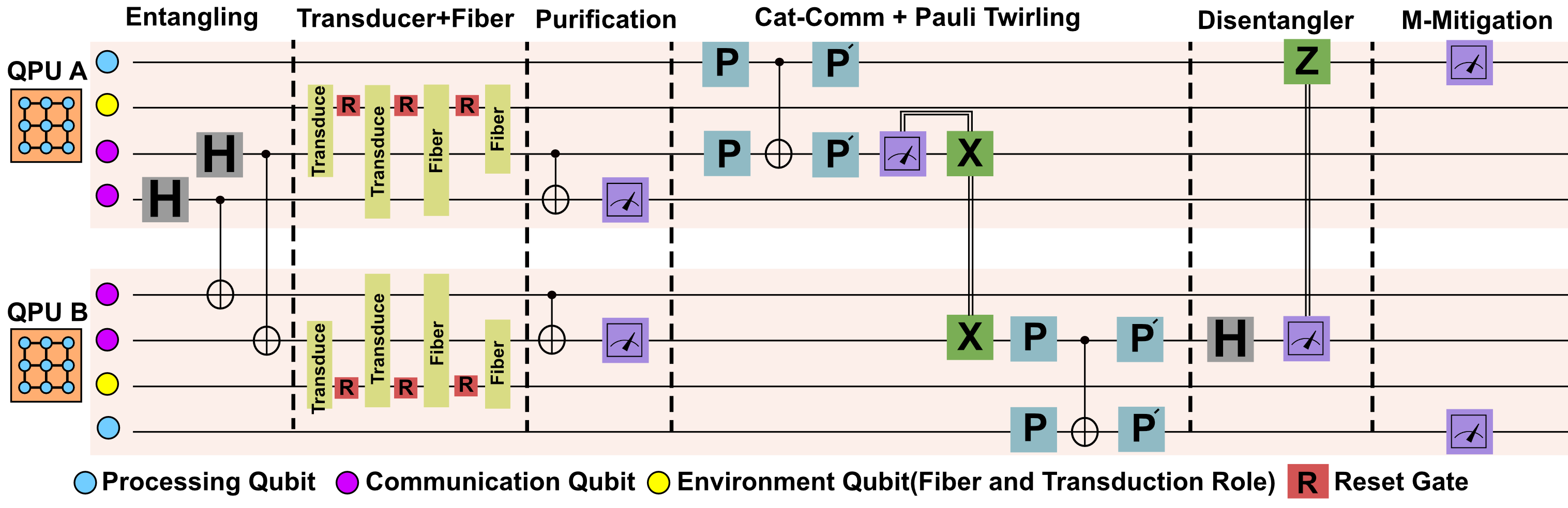}
\end{center}
\caption{Circuit-level implementation of the proposed error suppression framework. 
A Bell pair is distributed through a collisional communication channel modeling 
transduction and fiber noise. Entanglement purification is applied to improve 
resource fidelity, followed by remote gate execution using a cat-state protocol 
with Pauli-twirled two-qubit gates. Final measurement outcomes are corrected 
using readout mitigation.} 

\label{f:algorithm}
\end{figure*} 

We consider a distributed superconducting quantum computing architecture composed of two quantum processing units (QPUs) interconnected via a transduction interface and an optical fiber link, as shown in Fig.~\ref{f:Layout}(a). Each QPU comprises processing qubits (blue circles) for local operations and communication qubits (purple circles) for establishing nonlocal entanglement via flying qubits (green circles). Remote gates are realized by first distributing an entangled Bell pair between distant nodes 
and then consuming it through a remote gate protocol. 

To emulate communication-induced noise on current superconducting hardware, we adopt the quantum collisional model in which the environment is divided into smaller segments, each interacting with the system through a unitary time-evolution operator, as introduced in \cite{elyasi2025qdc}. As illustrated in Fig.~\ref{f:Layout}(a), the flying qubit sequentially interacts with environment ancilla qubits representing segments of the noisy channel, capturing both transduction and fiber propagation effects. Each interaction is described by the unitary
$
U_{SE} = e^{-i H_{SE}\Delta t},
$
with amplitude-damping Hamiltonian
$
H_{SE} = \kappa \left( \sigma^- \otimes \sigma^+ + \sigma^+ \otimes \sigma^- \right),
$
where $\kappa$ captures both transduction- and fiber-induced effects, with distinct coupling strengths for each process. After each interaction, the environment qubit is reset, yielding an effective Markovian channel. After $n$ collisions, the system evolves as
$
\rho_S^{(n)} = \mathrm{Tr}_{E} \left[ U_{SE}^{(n)} \cdots U_{SE}^{(1)} (\rho_S \otimes \rho_E) U_{SE}^{(1)\dagger} \cdots U_{SE}^{(n)\dagger} \right].
$
This approach lets us have controlled emulation of cumulative communication noise while remaining fully executable on monolithic hardware. 
It allows systematic tuning of transduction and fiber effects to evaluate their impact on entanglement distribution and remote gate fidelity.

\section{Error Suppression Model}

As shown in Fig.~\ref{f:Layout}(b) and Fig.~\ref{f:algorithm}, to improve robustness against communication-induced noise, we employ a layered error suppression strategy combining entanglement purification, Pauli twirling, and readout mitigation across different stages of the protocol.

First, as illustrated in Fig.~\ref{f:algorithm}, two Bell states $\ket{\phi^+}$ are generated using a Hadamard gate followed by a CNOT between communication qubits. These entangled pairs are then transmitted through the collisional channel (Transduction+Fiber), where sequential interactions via $U_{SE}$ operations—together with resetting the environment qubits at each step—emulate communication-induced noise under a Markovian assumption. This process captures both transduction and fiber effects prior to gate execution, resulting in degraded entangled resources that are subsequently used as input to the purification stage.

Entanglement purification ~\cite{dur1999} is implemented using a BBPSSW-type protocol ~\cite{bennett1996purification}, in which two imperfect Bell pairs are locally processed using local CNOT operations on both QPUs and subsequent measurement-based post-selection. 
Two noisy copies are processed via local operations and classical communication (LOCC), yielding
$
\rho' = \frac{\mathcal{P}(\rho \otimes \rho)\mathcal{P}^\dagger}{\mathrm{Tr}[\mathcal{P}(\rho \otimes \rho)]}.
$
This step suppresses a significant portion of the degradation introduced by transduction inefficiencies and fiber attenuation, increasing the usable entanglement fidelity at the cost of probabilistic success, and enabling more reliable remote gate execution.

At the gate level, the remote two-qubit operation is implemented using a cat-state-based protocol, in which the purified entangled resource is consumed via conditional operations, mid-circuit measurements, and feedforward corrections. To further improve robustness, Pauli twirling is applied to the unconditional two-qubit gates, transforming the effective noise channel as
$
\mathcal{E}_{\mathrm{twirled}}(\rho)
=
\frac{1}{4}
\sum_{P \in \{I,X,Y,Z\}}
P^{\dagger}\,
\mathcal{E}(P \rho P^{\dagger})\,
P.
$
Rather than eliminating noise, this process converts coherent errors into stochastic Pauli noise, thereby reducing their accumulation across the circuit and improving stability against calibration imperfections~\cite{wallman2016}.

Finally, readout mitigation is applied to correct bias in the measured output distributions~\cite{nation2021}. Using the assignment matrix $A$, with elements $A_{ij} = P(i|j)$, the corrected probabilities are obtained as
$
p_{\mathrm{ideal}} = A^{-1} p_{\mathrm{measured}}.
$
Since remote gate performance is evaluated directly from measurement outcomes, this step ensures that improvements achieved at earlier stages are accurately reflected in the final results.

\section{Results}

\begin{figure*}[t!]
\begin{center}
\includegraphics[width=1\textwidth]{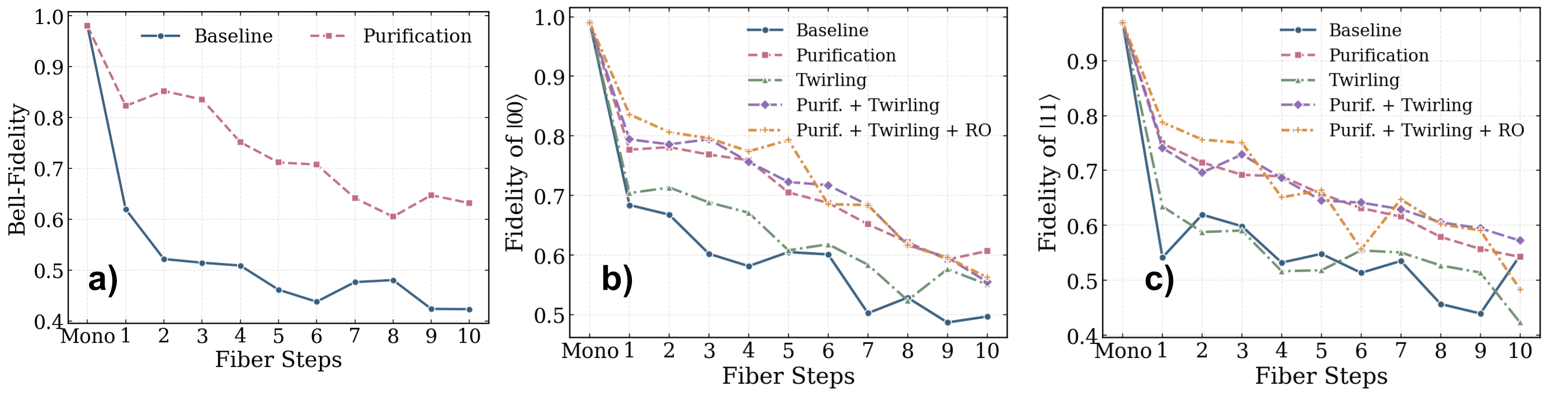}
\end{center}
\caption{Remote gate performance across increasing collisional-model fiber steps, with the monolithic noiseless case shown as Mono. (a) Quality of the distributed entanglement, quantified by the Bell-state fidelity. (b) Fidelity of $\lvert00\rangle$ for a remote CNOT with the control qubit prepared in $\lvert0\rangle$. (c) Fidelity of $\lvert11\rangle$ for control prepared in $\lvert1\rangle$.}
\label{f:results}
\end{figure*} 

Experiments are performed on the IBM \texttt{ibm\_kingston} backend, a 156-qubit Heron r2 superconducting quantum processor featuring tunable couplers and improved coherence properties compared to previous generations. All circuits are executed with 10,000 shots to ensure statistical reliability. Readout calibration matrices are obtained using standard measurement calibration circuits and inverted during post-processing to mitigate measurement bias.

Fig.~\ref{f:results} summarizes the performance of the remote gate protocol under increasing communication-induced noise, emulated via the quantum collisional model. The monolithic case (``Mono'') corresponds to an ideal local CNOT without transduction or fiber-induced effects, providing a reference fidelity of approximately $\sim 0.98$. As the number of fiber steps increases from 1 to 10, we map each interaction step to a physical separation using $D(n) = \frac{\gamma n}{\alpha}$, where $\gamma = \kappa^2$ denotes the coupling strength and $\alpha$ is the fiber attenuation constant. In this work, we emulate a standard G-652-D optical fiber with $\alpha = 0.0145~\mathrm{km}^{-1}$, where each step in the plots corresponds to a 10~m fiber segment. The cumulative interaction with the effective environment leads to a progressive degradation of both entanglement quality and remote gate fidelity.

Fig.~\ref{f:results}(a) presents the quality of the distributed entangled resource prior to gate execution, quantified using the Bell-state fidelity. In the baseline case, in which we do not employ any error suppression, the fidelity decreases rapidly from $\sim 0.97$ at Mono to $\sim 0.42$ at step 10, reflecting the strong impact of transduction inefficiencies and fiber attenuation on the shared entangled state. This trend is consistent with the accumulation of amplitude-exchange errors induced by the collisional model.

Applying entanglement purification significantly mitigates this degradation. The purified curves maintain fidelities of approximately $\sim 0.63$ at step 10, corresponding to an improvement exceeding $20\%$ in the high-noise regime. This demonstrates that purification effectively filters low-quality entangled pairs through probabilistic post-selection, thereby increasing the average fidelity of the remaining resource.

Figs.~\ref{f:results}(b) and (c) show the performance of the remote CNOT operation for control qubits initialized in $\lvert0\rangle$ and $\lvert1\rangle$, respectively. In both cases, the baseline implementation exhibits a monotonic decay from $\sim 0.97$ to below $\sim 0.50$ as the number of channel interactions increases, indicating that remote gate performance is directly limited by the degradation of the shared entangled resource, mainly caused by transduction and slightly by fiber attenuation.

Pauli twirling alone provides a moderate improvement, typically on the order of $\sim 5\%$ in intermediate regimes. This behavior arises from converting coherent gate errors into an effective stochastic Pauli channel, thereby reducing error accumulation across repeated operations. However, since twirling does not improve the underlying entanglement fidelity, its impact remains limited under strong communication noise.

The combination of entanglement purification and Pauli twirling leads to a better enhancement. In this case, success probabilities remain in the range of $\sim 0.58$--$0.62$ at step 10, demonstrating that improving both the resource quality and gate noise characteristics yields complementary benefits. The addition of readout mitigation (RO) further refines the observed results, providing the highest success probabilities across nearly all channel depths. Overall, the combined approach achieves improvements exceeding $20\%$ compared to the baseline in the transduction-limited regime.

Regarding the fidelity of  $\lvert00\rangle$ and  $\lvert11\rangle$,  we can observe that using Purification alone behaves closely to the case when it is applied together with Twirling and Readout mitigation, especially in the scenarios with more fiber steps (i.,e., higher noise levels). This is consistent with our previous findings that, in an interconnected paradigm, communication noise dominates accumulated error, making its mitigation crucial for the overall performance of the computing system. 


\section{Conclusion}

In this work, we present a practical approach to improve remote quantum gate fidelity in fiber-interconnected QPUs, incorporating communication noise mitigation through layered error suppression techniques. Results reveal that the combined use of entanglement purification, Pauli twirling, and readout mitigation effectively enhances the robustness of remote gate operations in interconnected settings. At the same time, for relatively long-distance interconnections, purification alone achieves fidelity comparable to that of the full combined approach. This work quantifies fidelity degradation in realistic distributed settings and supports the study of scalable and reliable quantum distributed-computing architectures.

\section{Acknowledgment}
This work is supported by the Swedish Research Council (VR), the Gender Initiative for Excellence (GENIE), and the UK EPSRC IQN Hub (EP/Z533208/1). We thank Chalmers Next Labs for providing a premium account and support for the IBM Quantum platforms.
\defbibnote{myprenote}{%
}
\printbibliography[prenote=myprenote]

\vspace{-4mm}

\end{document}